\documentclass[12pt]{article}
\usepackage{epsfig}

\def\beq{\begin{equation}}
\def\eeq#1{\label{#1}\end{equation}}
\def\eeqn{\end{equation}}

\def\beqa{\begin{eqnarray}}
\def\eeqa#1{\label{#1}\end{eqnarray}}
\def\eeqan{\end{eqnarray}}

\let\bar=\overbar

\def\Dslash{\not{\hbox{\kern-4pt $D$}}}
\def\dslash{\not{\hbox{\kern-2pt $\del$}}}

\def\msb{{\bar{\ssstyle M \kern -1pt S}}}

\def\Title#1{\begin{center} {\Large {\bf #1} } \end{center}}
\newcommand{\dedx}{d$E$/d$x$}
\newcommand{\pip}{$\pi^{+}$}
\newcommand{\pim}{$\pi^{-}$}
\newcommand{\kap}{$K^{+}$}
\newcommand{\kam}{$K^{-}$}
\newcommand{\pbar}{$\bar{p}$}

\begin{document}

\Title{Transverse momentum spectra of hadrons identified with the ALICE Inner Tracking System}

\bigskip\bigskip


\begin{raggedright}  

{\it Leonardo Milano on behalf of the ALICE Collaboration\index{Milano, L.}\\
Dipartimento di Fisica Sperimentale dell'Universit\`a \& Sezione INFN, Torino\\
10125 Turin, ITALY}
\bigskip\bigskip
\end{raggedright}

\section{Introduction}

The Inner Tracking System (ITS) is the ALICE detector closest to the beam axis. It is composed of six layers of silicon detectors: two innermost layers of Silicon Pixel Detectors (SPD), two intermediate layers of Silicon Drift Detectors (SDD) and two outermost layers of Silicon Strip Detectors (SSD). The ITS can be used as a standalone tracker in order to recover tracks that are not reconstructed by the Time Projection Chamber (TPC) and to reconstruct low momentum particles with $p_{\rm t}$ down to 100~MeV/$c$. Particle identification in the ITS is performed by measuring the energy-loss signal in the SDD and SSD layers. The ITS allows one to extend the charged particle identification capability in the ALICE central rapidity region at low $p_{\rm t}$: it is possible to separate $\pi/K$  in the range 100 MeV/$c$ $< p_{\rm t} <$ 500 MeV/$c$ and $K/p$ in the range 200 MeV/$c$ $ < p_{\rm t} <$ 800 MeV/$c$.

\section{PID technique}

\begin{figure}[htb]
\begin{center}
\begin{tabular}{cc}
\epsfig{file=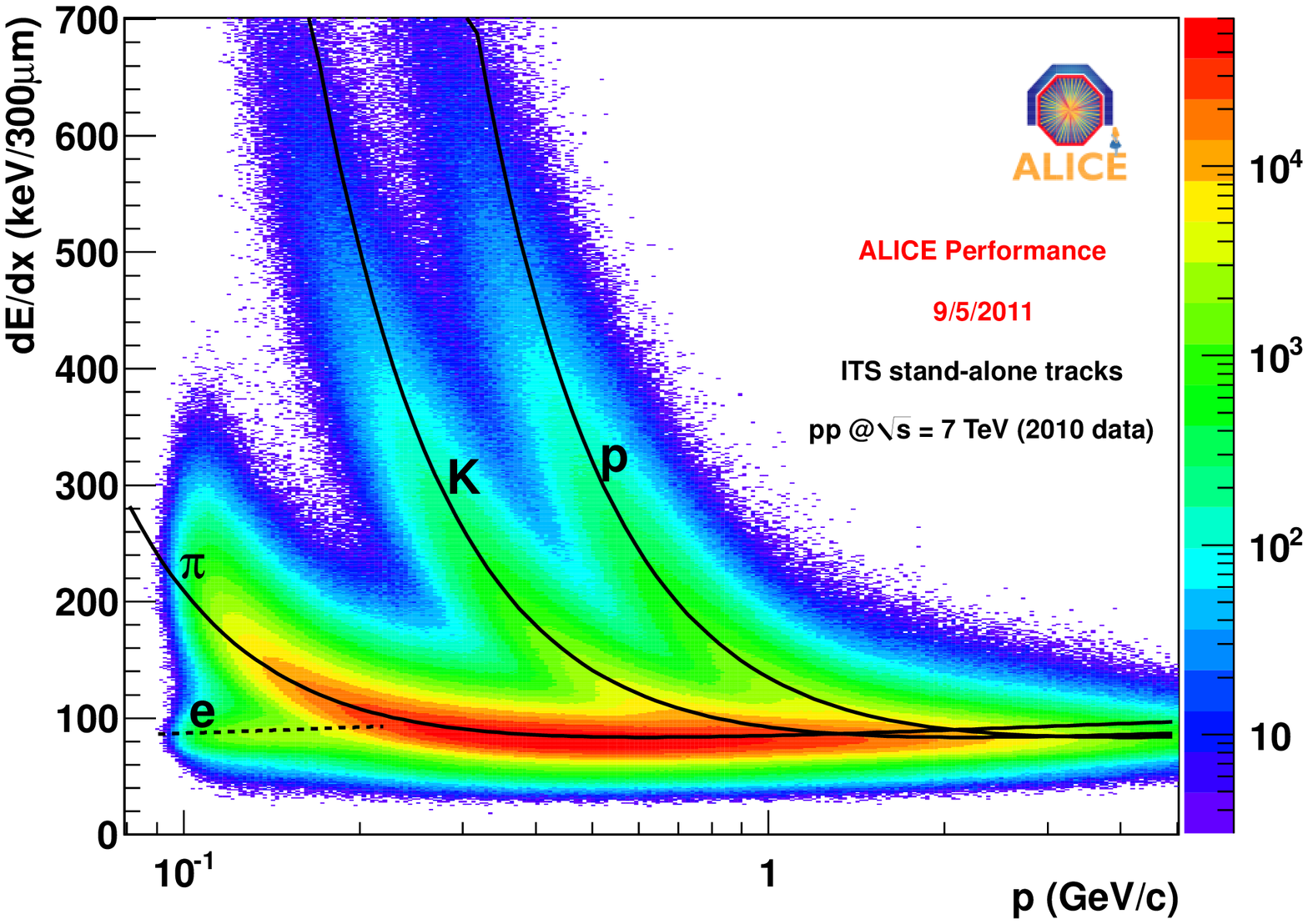,height=1.92in} &
\epsfig{file=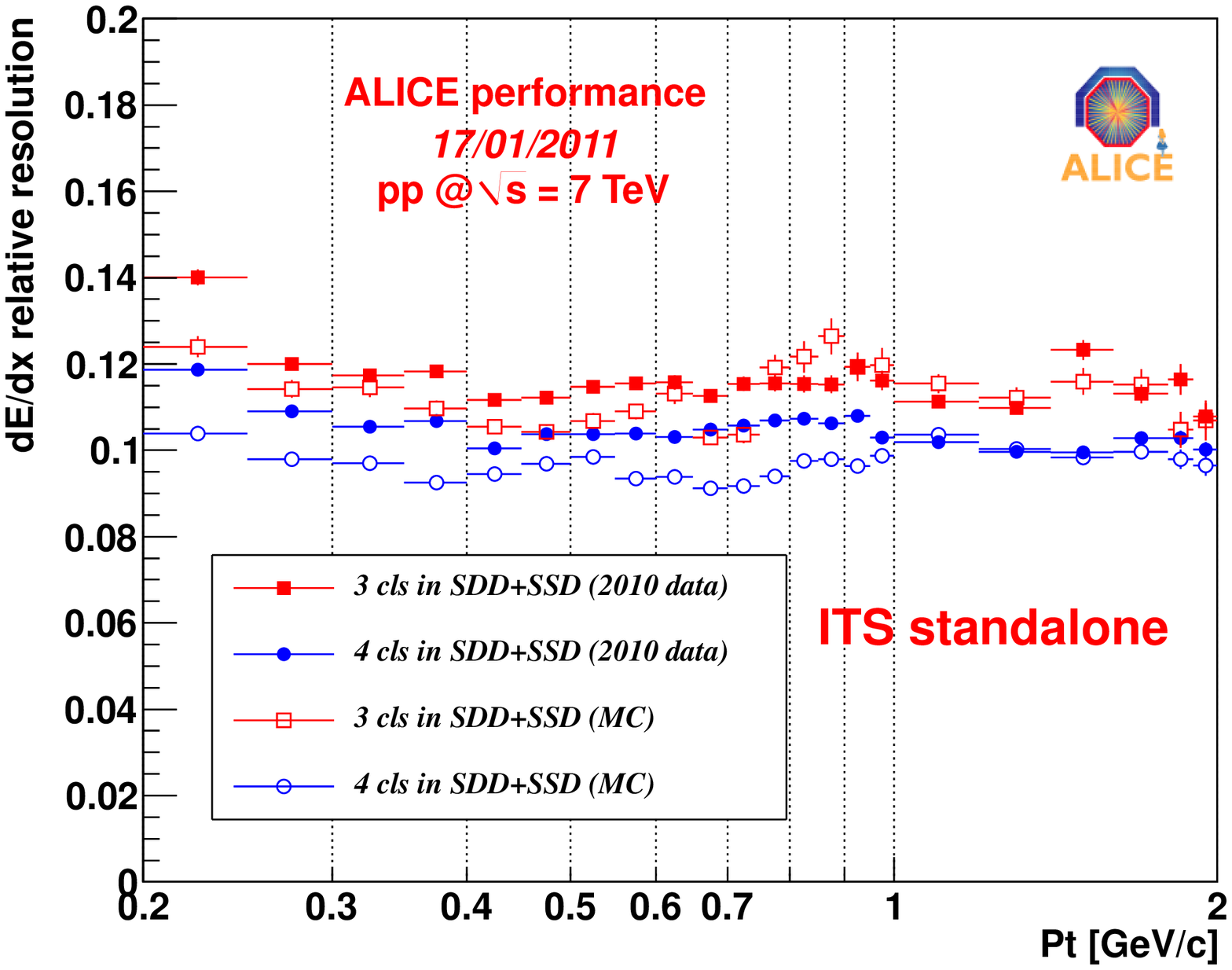,height=1.92in} \\
\end{tabular}
\caption{Specific energy-loss signal \dedx\ vs.~momentum (left) and resolution vs.~$p_{\rm t}$ (right) in pp collisions at $\sqrt{s}=$~7  TeV 
for ITS standalone tracks measured with the ITS. Solid lines in the left panel are a
parametrization (from \cite{Back:2006tt}).}
\label{fig:ITSsadedx}
\end{center}
\end{figure}

In both the ITS standalone (track reconstruction only using the ITS) and in the ITS-TPC (reconstruction performed using both the ITS and the TPC) analyses, the
\dedx\ measurement from the SDD and the SSD is used to identify
particles. The standalone tracking extends the momentum
range to lower $p_{\rm t}$ than can be measured in the TPC, while the
combined tracking provides a better momentum resolution. For each track, \dedx\ is calculated using a truncated mean: the average of
the lowest two points in case four points are measured, or a
weighted sum of the lowest (weight 1) and the second lowest point
(weight 1/2), in case only three points are measured. Figure~\ref{fig:ITSsadedx} shows the truncated mean \dedx\ for the
sample of ITS standalone tracks along with the PHOBOS
parametrization of the most probable value~\cite{Back:2006tt} as function of momentum and its resolution as a function of transverse momentum both for data and Monte Carlo simulation.

The raw hadron yields extracted from the fits to the \dedx\
distributions are corrected for the reconstruction efficiency
determined from Monte Carlo simulations, applying the same
analysis criteria to the simulated events  as to the data.
Secondary particles from interactions in the detector material and
strange particle decays are subtracted from the yield of
both simulated and real data.
 The secondary-to-primary ratio is estimated by fitting the
measured track impact-parameter distributions of each
hadron species with three components: prompt particles, secondaries from strange particle
decays and secondaries produced in the detector material. Alternatively, the contamination from secondaries
is determined using Monte Carlo samples, after rescaling
the $\Lambda$ yield to the measured values~\cite{strange}. The
difference between these two procedures is about 3\% for protons
and is negligible for other particles.


\section{Conclusions}

Particle Identification in the ITS allows one to extend the the ALICE PID capability at low $p_{\rm t}$.
The first analysis of transverse
momentum spectra of identified hadrons,  \pip, \pim, \kap, \kam,
$p$, \pbar, in pp collisions at  $\sqrt{s}=$~900  GeV  with the ALICE
detector is done, result are published in~\cite{pid}.
The measurements of these particle spectra is a substantial part of the ALICE program in both pp and PbPb collisions.


%
%
%
%
%

 
\end{document}